\documentclass[12pt]{article}
\usepackage{amsmath,amssymb,amsfonts,amsthm,graphicx}
\usepackage[bookmarksnumbered, colorlinks, plainpages]{hyperref}
\newcommand{\be}{\begin{equation}}
\newcommand{\ee}{\end{equation}}
\newcommand{\beq}{\begin{eqnarray}}
\newcommand{\eeq}{\end{eqnarray}}
\newtheorem{thm}{Theorem}
\newtheorem{lem}[thm]{Lemma}
\newenvironment{pf}{\par\noindent\textbf{Proof}.}{\hfill$\square$}
\title{A Time-space Trade-off for Computing the Geodesic Center of a Simple Polygon}
\author{\large Pardis Kavand$^1$, \hspace{.0cm}\large Ali Mohades $^1$\footnote{Corresponding author}, \hspace{.0cm} Mohammad Reza Kazemi$^1$\footnote{
{\em E-mail addresses:} kavandpardis@aut.ac.ir (P. Kavand), mohades@aut.ac.ir (A. Mohades), mr.kazemi@aut.ac.ir (M. Kazemi)$\newline$
} \hspace{.0cm} \vspace{.5cm}$ $\\
\small{$^{1}$ Laboratory of Algorithms and Computational Geometry,
Department of Mathematics and }\vspace{-1mm}\\
\small{ Computer Science, Amirkabir University of Technology.}}
\date{}

\begin{document}
 
\maketitle \thispagestyle{empty}\

\begin{abstract}
In this paper we study the problem of computing the geodesic center of a simple polygon when the available workspace is limited. For an n-vertex simple polygon, we give a time-space trade-off algorithm that finds the geodesic center in $O(T(n, s) \log^2 n + \frac{n^2}{s} \log n)$ expected time and uses $O(s)$ additional words of space where $s\in \Omega(\log n) \cap O(n)$, and $T(n, s)$ is the time needed for constructing the shortest path tree of a given point inside a simple polygon, in depth-first order, with $O(s)$ extra space. Applying the best current known time-space trade-off  of  Oh  and Ahn (Algorithmica 2019) for shortest path tree, our algorithm runs in $O(\frac{n^2}{s} \log^3 n)$ expected time.
\\
\\
$\mathbf{Keywords}$: computational geometry, memory-constrained algorithms, constrained geodesic center, geodesic center.
\end{abstract}
\maketitle
\section{Introduction}
\label{sec:1}
This paper considers a  long-standing problem known as the geodesic center problem, which involves finding a point inside a given simple polygon $P$ of $n$ vertices, that minimizes the maximum geodesic distance to any point inside $P$.  Pollack et al. showed that the problem can be solved in $O(n \log n)$ time~\cite{14}. For a long time, it was the best result for the problem. Finally, in a fairly recent paper, Ahn et al. provided a linear time algorithm for the problem~\cite{29}. Both algorithms use $O(n)$ amount of space. We are interested in studying the problem when the available workspace is limited (specifically sub-linear). 
Our underlying memory-constrained model is as follows: the input of the problem is given in a read-only memory, such that random access to each input item is possible. In addition to the input, we are permitted to use $O(s)$ words of space for reading and writing, for any $s \in \{1, \dots, n\}$. Note that, a word of space has $\Theta(\log n)$ bits. Since, the size of the intermediate data can be larger than the available workspace,  they are not stored but computed when needed. In literature, this often referred to as the $s$-workspace model. 
There are several algorithmic results in this setting in computational geometry. In all of them, either a new algorithm is devised, or an existed algorithm in unlimited environment is adapted to the memory-constrained setting. For a survey of memory-constrained algorithms in computational geometry, the reader is referred to~\cite{30}. 

A problem that has been considered in this model is the problem of computing the shortest path between two given points inside a simple polygon. Har-Peled showed that there exists an $s$-workspace algorithm that computes the shortest path between two points inside a simple polygon of $n$ vertices in $O(n^2/s + n\log s\log^4(n/s))$ expected time~\cite{31}.
Aronov et al. presented an algorithm for constructing the shortest path tree of a given point inside an $n$-vertex simple polygon. Their algorithm runs in $O(\frac{n^2}{s}\log n + n \log s \log^5 (n/s))$ expected time and uses $O(s)$ words of space, for any $s\leq n$~\cite{4}. Oh and Ahn improved these results~\cite{27}. Both algorithms are $s$-workspace algorithms. Their algorithm for computing the shortest path runs in $O(n^2/s)$ deterministic time. The expected running time of their algorithm for constructing the shortest path tree is $O(\frac{n^2}{s} \log n)$.

We present an $s$-workspace algorithm that finds the geodesic center of a simple polygon $P$ of size $n$ in $O(T(n, s) \log^2 n + \frac{n^2}{s} \log n)$ expected time, assuming $s \in \Omega(\log n) \cap O(n)$, where $T(n, s)$ is the time complexity of constructing the shortest path tree of a given point inside $P$, in depth-first order, using $O(s)$ words of space. Our approach provides a time-space trade-off; that is, the running time of our algorithm decreases as $s$ grows. If we use the shortest path tree algorithm of Oh and Ahn~\cite{27}, the expected running time of our algorithm will be $O(\frac{n^2}{s}\log^3 n)$.

The rest of this paper is organized as follows: Section 2 contains some preliminaries and notations. In Section 3, we study the problem of computing the geodesic center of a given simple polygon. To solve the problem, we need to know how to find the geodesic center constrained to a chord of the polygon. Thus, we first consider the constrained geodesic center problem (Section 3.1). Afterwards, in Section 3.2, we explain how to solve the geodesic center problem. Finally, we give a conclusion in Section 4.

\section{Preliminaries and Notations}
\label{sec:2}
Assume that $P$ is a simple polygon of $n$ vertices, and $p$ and $q$ are two points inside $P$. The \textit{shortest path} from $p$ to $q$ in $P$ is a polygonal path connecting $p$ to $q$ with minimum length that is totally contained in $P$. 
The \textit{geodesic distance} between $p$ and $q$, denoted by $d_P(p, q)$, is defined as the length of the shortest path between $p$ and $q$ in $P$. 
The \textit{direction} $\vec{v}(p, q)$ is the unit vector at $p$ directed along the first segment of the shortest path from $p$ to $q$ in $P$.  
A \textit{geodesic farthest neighbor} of $p$ in $P$ is a point of $P$ with maximum geodesic distance from $p$. It is well-known that a geodesic farthest neighbor is always a vertex of the polygon~\cite{14}. Also, a point can have more than one geodesic farthest neighbor. We denote by $F_P(p)$ the set of all geodesic farthest neighbors of $p$ in $P$.

The \textit{shortest path tree} of $p$ inside $P$ is a tree rooted at $p$ whose nodes represent all vertices of $P$, and there exists an edge between two nodes $u$ and $v$ if and only if the line segment $uv$ lies on either $\pi_P(p, u)$ or $\pi_P(p, v)$. We use $ST_P(p)$ to denote the shortest path tree.

A \textit{chord} of $P$ is a line segment whose endpoints lie on the boundary of $P$ and does not intersect the exterior of $P$.

Recall that, throughout the paper, by $T(n, s)$ we mean the time needed for constructing the shortest path tree of a given point inside $P$, in depth-first order, with $O(s)$ extra space.

\section{Geodesic Centers}
\label{sec:3}
In this section we propose an $s$-workspace algorithm which computes the geodesic center of a given simple polygon $P$ of $n$ vertices in $O(T(n, s) \log^2 n + \frac{n^2}{s} \log n)$ expected time, where $s\in \Omega(\log n) \cap O(n)$. Our approach is similar to that of Pollack et al.~\cite{14}. They showed that there exists an algorithm that finds the geodesic center of a simple polygon with $n$ vertices in $O(n \log n)$ time using $O(n)$ space. We apply their approach to solve the geodesic center problem in the aforementioned model of memory-constrained. 

Before going into the details of our algorithm, let us first briefly explain the approach. Initially we show, for a given chord inside $P$, how to decide on which side of the chord the geodesic center lies. To do that, we find the geodesic center constrained to the chord. Using these and performing a binary search, we will find a triangle inside the polygon which contains the geodesic center. 

To find the geodesic center, constrained to a chord or a triangle, we first compute the shortest path tree of the unknown center. Having the shortest path tree of the center in hand, we can find a linear number of circles such that finding the geodesic center reduces to computing the smallest circle enclosing these circles. 

Here, we show how to do all these in the memory-constrained environment. Thus, in Subsection 3.1, we consider the constrained geodesic center problem. Then, in Subsection 3.2, we explain how to find the geodesic center of $P$.

\subsection{Constrained Geodesic Centers}
We are given a simple polygon $P$ and a chord $d$ of $P$. We show how to find the geodesic center constrained to $d$. In other words, we present an algorithm for computing a point $c_d$ of $d$ which minimizes the maximum geodesic distance to any point inside $P$. Our algorithm is a memory-constrained algorithm that runs in $O(T(n,s)\log n)$ expected time and uses $O(s)$ words of space for any $s\in \Omega(\log n)\cap O(n)$. First, we briefly describe our algorithm, then we go into more details.
\\[2mm]
\noindent\textbf{Algorithm.} Let $a$ and $b$ be the endpoints of $d$. Consider all edges of  $ST_P(a)$ and $ST_P(b)$ and the intersections of their extensions with $d$, regardless of whether the extensions meet the boundary of $P$. Let $X$ be the sequence of the intersection points, in the order that they are appear on $d$, and $m$ be the number of the members of the sequence. $X$ divides $d$ into $m+1$ subchords in which the shortest path tree does not change. 
In other words, for any such subchord $I$ and for each vertex $p$, there exists a reflex vertex $r_p$ such that the shortest path from any point $x\in I$ to $p$ is of the form  $xr_p \cup \pi_P(r_p, p)$, 
in which $xr_p$ represents the line segment connecting $x$ to $r_p$. 
By performing a binary search on $X$, 
we can identify the subchord containing the constrained geodesic center (Lemma~\ref{lem:2}).  
The problem of computing the constrained geodesic center in the subchord is precisely equivalent to that of computing the smallest circle enclosing all circles centered at $r_p$ with radius $d_P(r_p, p)$ for each vertex  $p$ of $P$.
To solve the problem, we use an approach similar to the method used by Megiddo~\cite{26}. We modify the method in such a way that it uses sublinear space (Lemma~\ref{lem:7}).

\begin{lem}
\label{lem:2}
We can compute the shortest path tree of the unknown constrained geodesic center in $O(T(n, s)\log n)$ expected time using $O(s)$ additional words of space for any $s \in O(n)$.
\end{lem}

\begin{pf}
From above discussion, to compute the shortest path tree of the constrained geodesic center, it suffices to find the subchord where contains it. To this end, we perform a binary search on $X$.
At the end of the $i$th step of the search, we find a subchord $I_i$ of $d$ of size at most $(\textstyle\frac{3}{4})^im$ which contains the constrained geodesic center. By \textit{size of a subshord} we mean the number of the intersection points which are contained in the subchord. 
Initially, we set $I_0$ to be the whole $d$.  While generating the edges of the shortest path trees, we randomly pick  points from $X$ until we finally get close enough to its median. Note that we do not need to store $X$ in the space, instead, whenever we need it,  we generate it using the shortest path tree algorithm.  The expected number of the iterations to find such a point is constant.
In other words, after constant number of iterations, we expect to detect a point $x_1 \in X$ that splits $I_0$ into two subchords of size at most $\frac {3}4m$ each. 
By constructing $ST_P(x_1)$, we compute the geodesic farthest neighbor(s) of $x_1$ in $P$. 
Since the constrained geodesic center problem is a convex programming~\cite{14}, 
if the directions $\{\vec{v}(x_1, f) \vert f\in F_P(x_1)\}$ lie on different sides of the line perpendicular to $d$ at $x_1$, then $x_1$ is the constrained geodesic center. Otherwise, the center lies on the same side as the directions. 
In this case, we consider the subchord containing the center as $I_1$, and ignore the other subchord. 
In this way, at the beginning of the $(i+1)$th step, we have a subchord $I_{i}$ of $d$ of size at most $(\textstyle\frac{3}{4})^{i}m$ which contains the constrained geodesic center. 
Again we construct $ST_P(a)$ and $ST_P(b)$. Using a similar approach as above on the intersection points that are contained in $I_{i}$, we find a point $x_{i} \in X$ that splits $I_{i}$ into two subchords of nearly equal size, and compute the geodesic farthest neighbor(s) of $x_{i}$. Thus, at the end of the step, we will find the subchord $I_{i+1}$ of size at most $(\textstyle\frac{3}{4})^{i+1}m$ where the constrained geodesic center lies. 
Since shortest path trees are of linear size, after $O(\log m)$ steps, we will identify the subchord where contains the constrained geodesic center but no intersection point. As a result, the shortest path tree does not change in the subchord. Therefore, the shortest path tree of the constrained geodesic center is same as that of any point in the subchord. Note that in each step, the expected number of times we call the  shortest path tree algorithm is constant. Therefore the result follows.
\end{pf}

Using above lemma, we can find the subchord $I^*$ of $d$ containing the constrained geodesic center with this property that the shortest path tree does not change in it. 
Let $R^*$ be the set of all nodes that are children of the root in the shortest path tree $ST_P(x)$, for any point $x\in I^*$.
For each $r\in R^*$, we denote by $f_r$ the geodesic distance between $r$ and its geodesic farthest neighbor(s) in the subtree rooted at $r$ in $ST_P(x)$.
As we mentioned before, the problem of computing the constrained geodesic center in $I^*$ is equivalent to that of computing the smallest circle enclosing circles centered at $r$ with radius $f_r$, for all $r\in R^*$. 

To solve the problem, we apply the prune-and-search algorithm by Megiddo for computing the smallest enclosing circle of pints in the plane~\cite{26}. We use \textit{tournament tree} to adapt the algorithm in such a way that it can be used in the memory-constrained model. Hence, we first explain the Megiddo's approach by using tournament tree. Next, we show how to modify it to solve our problem in the memory-constrained model.

We are given a line $l$ and a set $S$ of $n$ points in the plane. We aim to find the smallest circle enclosing $S$ whose center is constrained to lie on $l$. 
Megiddo's algorithm for solving the problem is a prune-and-search algorithm with $O(\log n)$ rounds, in which, in each round, $1/4$ of the points of $S$ are pruned. At each round, the set of all points that have not been pruned in previous rounds are called \textit{active points}. 
At the beginning of each round, the algorithm pairs the active points, intersects their bisectors with $l$, finds the median of the intersection points, and decides on which side of the median the solution lies. 
Next, for each pair whose corresponding bisector intersect $l$ at the side of the median that does not contain the solution, the algorithm prunes the point that is closer to the median. In this way, at the end of the round, $1/4$ of the active points are omitted. 
Thus, after $O(\log n)$ rounds, we have constant number of active points whose smallest enclosing circle can be found directly.

We can explain the process using tournament tree. In fact, we have a tournament tree with \textit{draw possibility}. We number the levels of the tree from bottom to top. All points of $S$ are represented by the leaves of the tree. The first round constructs the second level of the tree.  The $i$th round of the algorithm constructs $(i+1)$th level of the tree using the $i$th level. We build the tournament tree from bottom to top. Nodes in the $i$th level of the tree represent the active points in its corresponding round. 
The number of nodes in $(i+1)$th level of the tree is at most $3/4$ of that in $i$th level.  
 At the beginning of the $i$th round, we pair nodes of the $i$th level. 
 For pairs one of whose points is pruned, we call the pruned point \textit{loser} and the other one \textit{winner}. Winners are qualified to advance to the $(i+1)$th level. For other pairs, we say that \textit{draw} happens. In this case, both points will qualify to advance to the $(i+1)$th level.

Now, suppose that a limited amount of workspace is available. We aim to find the smallest circle enclosing circles $C(r, f_r)$, for all $r\in R^*$, such that the center of the circle lies on the subchord $I^*$, where $C(r, f_r)$ denotes the circle centered at $r$ with radius $f_r$. We use a similar approach as above.
Note that, to construct each level of the tournament tree, we need to know the lower level. Since the available workspace is limited, we are not able to store lower levels, instead, at each level, we construct all lower levels parallelly.

Suppose that we are at the beginning of the first round and constructing the second level of the tournament tree. We have a subchord $I^*$ where we are looking for the solution. We construct $ST_P(x)$ for an arbitrary point $x$ inside $I^*$. 
Let $C_1=C(r_1, f_1), \dots, C_{n^\prime}=C(r_{n^{\prime}}, f_{n^\prime})$ be the sequence of the circles that are going to be enclosed, in the order that they are generated while constructing $ST_P(x)$. 
We pair $C_{2j-1}$ and $C_{2j}$ for each $j\in \{1, 2, \dots, {n^{\prime}}/2\}$. We define \textit{covering bisector} of $C_{2j-1}$ and $C_{2j}$ to be the locus of all points that are equidistant from the farthest  points on $C_{2j-1}$ and $C_{2j}$. Note that the covering bisectors are hyperbolic, thus they intersect a line in at most two points. 
For any pair whose covering bisector do not intersect $I^*$, one circle can be ignored (the one that lies at the same side of the covering bisector as $I^*$), because it is covered if so is the other one.
Let $T$ be the sequence of all intersection points of the covering bisectors of $C_{2j-1}$ and $C_{2j}$ with $I^*$, for each pair intersecting $I^*$, in the order that they appear on $I^*$.
By computing the approximate median of $T$, in a similar way as the proof of Lemma~\ref{lem:2}, we can find a subinterval of $I^*$ containing  at least $1/4$ of the intersection points but not the solution. For the covering bisectors that intersect this subinterval, we compute the approximate median of the intersection points that lie in the subinterval containing the solution. In this way, we will find a subinterval of $I^*$ that is not intersected by at least $1/{16}$ of the covering bisectors. We use $J_1$ to denote the subinterval.
For pairs whose covering bisector does not intersect $J_1$, with the same argument as above, we can ignore one of the circles. In this case, only one of the circles is active at the higher level. For other pairs draw happens.

For integer $i \geq 1$, we let $A_i$ to be a process that computes an interval $J_i$ at the $i$th level which contains the solution and at least $1/{16}$ of the covering bisectors of the active circles do not intersect it. We also define $B_i$ to be a process that constructs the $(i+1)$th level of the tree. We described $A_1$ and $B_1$ at the previous paragraph. 
For $i \geq 2$, $A_{i}$ computes $J_{i}$ by running $B_{i-1}$ constant times. 
We maintains $J_{i}$ to be used by $B_{i}$. $B_{i}$ uses $J_{i}$ and runs $B_{i-1}$ once to construct the $(i+1)$th level.
Everything is similar to the previous paragraph, except that we get pairs by using $B_i$ instead of the shortest path tree algorithm.

Let $\tau_i$ and $s_i$  be the running time and workspace needed for constructing $(i+1)$th level, respectively. From above discussion, $\tau_{i}=\tau_{i-1}+O(n/\sigma^{i-1})=\tau_1+O(n)$ and $s_{i} = s_{i-1} +c=ic$, where $c$ and $\sigma \geq \frac{32}{31}$ are constants. Also, the time for computing $J_i$ is $c'\tau_{i-1}+O(n/\sigma^{i-1})=c'\tau_1 +O(n)$, for a constant $c'$. 

 In this way, after constructing $O(\log n)$ levels, we have constant number of active circles. Therefore, we can find the constrained geodesic center directly. The following lemma summarizes the result.

\begin{lem}
\label{lem:7}
Let $P$ be a simple polygon with $n$ vertices, and $d$ be a chord of the polygon. We can compute the geodesic center constrained to $d$ in $O(T(n,s)\log n)$ expected time and $O(s)$ extra space, where $s\in \Omega(\log n) \cap O(n)$. 
\end{lem} 

\begin{pf}
Since at each level of the tournament tree a constant fraction of nodes are dominated, the tree has $O(\log n)$ levels. 
Therefore, from above discussions, the total time spent by our algorithm is given by $O(\log n)((c'+1)\tau_1 +O(n))$. $B_1$ constructs the shortest path tree constant times. Hence, $\tau_1=O(T(n,s))$. Note that $i\in O(\log n)$ and $s_i = ic$, for some constant $c$. Thus, in addition to the space needed for constructing the shortest path tree, we use $O(\log n)$ words of space. As a result the time and space bounds follow. The correctness of our algorithm is deduced from that of Megiddo's.
\end{pf}

\subsection{Geodesic Centers}

In this section, we aim to show how to find the geodesic center of a simple polygon $P$ in the memory-constrained setting. We first find a chord splitting $P$ into two subpolygons of almost equal size (Lemma~\ref{lem:3}). Then, we decide on which side of the chord the geodesic center lies, and repeat the process for the subpolygon recursively. At the end, we will identify a triangle whose edges are chords of $P$ and contains the geodesic center (Lemma~\ref{lem:5}). Afterwards, we find a subregion of the triangle containing the geodesic center where the shortest path tree does not change. In this way, we can compute the shortest path tree of the geodesic center. As we mentioned before, having the shortest path tree in hand, we can find a linear number of circles such that finding the geodesic center reduces to computing the smallest circle enclosing these circles. We will show how to take care of this in the memory-constrained model.

There are some algorithmic results in the memory-constrained environment for decomposing simple polygons into subpolygons. To our knowledge, the best one presents an $O(n^ 2 /s)$-time $s$-workspace algorithm for subdividing a simple polygon of $n$ vertices into $O(\min\{n/s, s\})$ subpolygons of complexity $O(\max\{n/s, s\})$~\cite{27}. Note that, it suffices for us to partition $P$ into only two almost equal-sized subpolygons. Hence, we can conclude the following lemma from Lemma~1 in \cite{27}.

\begin{lem}
\label{lem:3}
We can find in $O(n^2/s)$ time a chord that decomposes $P$ into two nearly equal-sized subpolygons. Our approach uses $O(s)$ words of space. 
\end{lem}

\begin{pf}
There exists a vertical chord $l$ of $P$ that splits $P$ into two subpolygons of size at most $\frac{2}{3}n$ each~\cite{24}. By moving $l$ horizontally in $P$ such that it is a chord yet, it will touch a vertex of $P$. Hence, it suffices to check for each vertex $v$ of $P$ whether the vertical chord passing through $v$ partitions $P$ into two nearly equal-sized parts. Lemma~1 in \cite{27} explains how to find vertical chords passing through all vertices of a simple polygon in $O(n^2/s)$ time using $O(s)$ extra space. Therefore, the result follows.
\end{pf}

\begin{lem}
\label{lem:5}
Let $P$ be a simple polygon of $n$ vertices. We can find the geodesic center of the polygon or a triangle whose edges are chords of $P$ and contains the geodesic center in $O(T(n,s)\log^2 n + \frac{n^2}{s} \log n )$ expected time and $O(s)$ words of space where $s \in \Omega(\log n) \cap O(n)$.
\end{lem}

\begin{pf}
Using Lemma~\ref{lem:3}, we first find a chord that partitions $P$ into two nearly equal-sized subpolygons. Next, we decide in which subpolygon the geodesic center lies and repeat the process on the subpolygon. In each step, we have a subpolygon that contains the geodesic center of $P$. We call the subpolygon the \textit{underlying subpolygon} of the step. $P_i$ is used to denote the underlying subpolygon of the $i$th step. Let $d_i$ be the chord that splits $P_i$ into two subpolygons of almost equal size.
$d_i$ also divides $P$ into two subpolygons $P_l^i$ and $P_r^i$ (Note that these chords $d_i$ are vertical and do not intersect each other). We find the  geodesic center of $P$ constrained to $d_i$. We use $c_i$ to denote the center. By constructing the shortest path trees of $c_i$ in $P$, we can compute its geodesic farthest neighbors  in $P$. Let $\vec{v}_p(q)$ be the unit vector from $p$ in direction of the first edge on $\pi_P(p, q)$, and $F_P(p)$ denotes the set of all geodesic farthest neighbors of $p$ in $P$. From Lemmas 2 and 3 in \cite{14}, if $\{\vec{v}_{c_i}(f)\vert f \in F_P(c_i)\}$ does not lie in an open halfplane through $c_i$, then $c_i$ is the geodesic center of $P$. Otherwise, the unit vector that bisects the angle of the cone with apex $c_i$ spanned by all directions of $\{\vec{v}_{c_i}(f)\vert f \in F_P(c_i)\}$, points to the side of $d_i$ containing the geodesic center. Note that, since a cone can be represented by only two vectors, we do not need to maintain all these vectors. 
Assume without loss of generality that $P_l^i$ contains the geodesic center, and we set $P_{i+1} = P_i \cap P_l^i$ and consider $P_{i+1}$ as new underlying subpolygon. 
We repeat the process until the underlying subpolygon is of constant size. Now by triangulating the subpolyon and finding the constrained geodesic center for the diagonals, we will eventually find the geodesic center or  a triangle that contains the geodesic center. 

The above process has $O(\log n)$ steps. 
In each step, we compute a chord and the side where the geodesic center lies. The underlying subpolygon of each step can be determined by $O(\log n)$ chords and directions from previous steps. Using these and by applying Lemma~\ref{lem:3}, we can construct the underlying subpolygon of the next step. Thus, from Lemmas~\ref{lem:7} and \ref{lem:3} and the time and space bounds of the shortest path tree algorithm the result follows.
\end{pf}

Let $\bigtriangleup{\alpha\beta\gamma}$ be the triangle from Lemma~\ref{lem:5}. We now show how to find the geodesic center inside the triangle. Similar to the constrained case, we find a subregion of the triangle which contains the geodesic center, additionally, the shortest path tree does not change in it. To do that, we perform an approach similar to~\cite{14} and adapt it to be able to be used in the aforementioned model of memory-constrained.    

Our algorithm has $O(\log n)$ rounds. In each round, two half-planes containing the geodesic center are identified. Therefore, after $O(\log n)$ rounds, we have a set $L^\star$ of $O(\log n)$ half-planes defining a subregion in $\bigtriangleup{\alpha\beta\gamma}$ which contains the geodesic center, with this property that the shortest path tree does not change in the subregion. 

We denote by $L$ the set of lines passing through all edges of the shortest path trees $ST_P(\alpha)$, $ST_P(\beta)$ and $ST_P(\gamma)$.
At the beginning of the first round, we set $L^\star$ to be an empty set. Using random choices similar to that of the proof of Lemma~\ref{lem:2}, we find a line with approximately median slope among the members of $L$. Let $l_m^1$ denotes the line. We rotate all lines in $L$ around an arbitrary point on $l_m^1$, considering $l_m^1$ as $x$-axis. 
Next, we pair the lines such that in each pair, one line has positive slope and the other has negative slope. To do that, we run the shortest path tree algorithm twice for each of $ST_P(\alpha)$, $ST_P(\beta)$ and $ST_P(\gamma)$, in parallel fashion, such that one reports the positive-slope lines in $L$ and the other reports the negative-slope ones. Now, we find the median of $x$-coordinates of the intersection points of lines in pairs. We use $x_m^1$ to denote the median. By solving the constrained geodesic center problem for $x=x_m^1$, we can decide on which side of $x=x_m^1$ the geodesic center lies. 
We now find the median of $y$-coordinates of the intersection points lying on the opposite side. Let $y_m^1$ denotes the median. We now find the geodesic center constrained to $y=y_m^1$ to determine the side on which the geodesic center lies. 
We add to $L^\star$ two half-planes $h_x^1$ and $h_y^1$ defined by $x=x_m^1$ and $y=y_m^1$, respectively, and contain the geodesic center. 
Note that, at least $1/{16}$ of pairs do not have their intersection point inside $h_x^1\cup h_y^1$. For each such pair, the line which does not intersect $h_x^1\cap h_y^1$ can be ignored at the next round. Other lines are called \textit{active} at the next round.
As a result, in each round, a constant fraction of active lines are omitted.

In this way, after $O(\log n)$ rounds, we have $O(\log n)$ half-planes defining a subregion in $\bigtriangleup{\alpha\beta\gamma}$ which contains the geodesic center. Furthermore, the number of lines in $L$ intersecting the subregion is constant. 
For these lines, we solve the constrained geodesic center problem directly to determine on which side of them the geodesic center lies. We add to $L^*$ their corresponding half-planes. 
In this way, we will construct a subregion of $\bigtriangleup{\alpha\beta\gamma}$ whose interior is not intersected by any lines in $L$. 
This means that the shortest path tree does not change in the subregion. In other words, if $R$ denotes the subregion, for any two points $u, v \in R$, the shortest path trees $ST_P(u)$ and $ST_P(v)$ are same.
Let $x$ be an arbitrary point in $R$, points $q_1, \dots, q_k$ be the vertices of $P$ which form the first level of $ST_P(x)$, and $f_i$ be the distance between $q_i$ and its farthest neighbor(s) in subtree of $ST_P(x)$ rooted at $q_i$, for each $i = 1, \dots, k$.  
The problem of computing the geodesic center of $P$ is equivalent to that of finding the center of the smallest circle that encloses circles $C(q_1, f_1), \dots, C(q_k, f_k)$. If $C(x, \rho)$ denotes the circle, we can state the problem as fallows:
\begin{eqnarray}
\label{eq1}
\mathrm{Minimize}       &&    \rho \\ \nonumber
  \mathrm{subject~to}  &&   \|x - q_i\| + f_i \leq \rho \quad (1\leq i\leq k).
\end{eqnarray}

This is an optimization problem in three dimensions that can be solved in linear time when there is no limitation on the space. We use the approaches presented in \cite{22} and \cite{28} to solve the problem in the memory-constrained environment. 

Although the constraints are not linear, according to \cite{22}, for each pair of the constraints, there exists a plane such that if we know the position of the solution relative to the plane, we can omit one of the constraints.
Suppose that we have an oracle that can find the position of the solution relative to a given plane. Using Meggido's approach~\cite{28}, we want to solve the problem by calling the oracle a few number of times. 
Meggido's approach is a prune-and-search method that in each step produces constant number of half-spaces. 
Using these half-spaces, it drops a constant fraction of planes. In fact, in each step, it has constant number of median finding and plane pairing, alternatively. Accordingly, it finds constant number of planes. By applying the oracle for the planes, it obtains the mentioned half-spaces. 
To find median, we run the shortest path tree algorithm constant expected number of times. To pair planes, we run the shortest path tree algorithm two more times, parallelly, one for generating positive slopes and the other for negative ones. Later, we will explain how to implement the oracle in our computational model.

Now let us return to the approach explained in the previous section. Similarly, the process $A_i$ computes a subspace containing the solution. Using this, the process $B_i$ reduces the size of the set of active planes by a constant fraction and determines the set of the active planes in higher level. Note that $A_i$ is nothing but what explained in the previous paragraph. The intended subspace is the intersection of the half-spaces.
In fact, $A_i$ runs the shortest path tree algorithm and calls the oracle constant number of times. Consequently, its running time is from the order of the maximum of the running time of the oracle and the shortest path tree algorithm. Hence, everything is same as before except that we need to explain the oracle.

The oracle is an algorithm that for the problem~\ref{eq1} and a given plane $h$, decides whether the solution lies on $h$, or else which of the half-spaces bounded by $h$ contains the solution in its interior. To do that, it first solves the problem~\ref{eq1} constrained to $h$. 
Next, it finds the direction that decreases the cost function of the unconstrained problem. 
Note that the constrained problem is nothing but the unconstrained problem at a lower dimension. Therefore, it can be solved in recursive manner. The base case of the recursion is very similar to computing the geodesic center constrained to a line. Since we recurse on dimension, the running time of the oracle is from the order of that of the constrained geodesic center algorithm (See Lemma~\ref{lem:7}). Thus, we have the following result.

\begin{thm}
\label{theo:8}
There is an $s$-workspace algorithm that computes the geodesic center of a simple polygon with $n$ vertices in $O(T(n, s) \log ^2 n + \frac{n^2}{s} \log n)$ expected time, where $s\in\Omega(\log n) \cap O(n)$.
\end{thm}

\section{Conclusion}

In this paper we addressed the problem of computing the geodesic center of a simple polygon. For an $n$-vertex simple polygon $P$, we provided a time-space trade-off algorithm that solves the problem in $O(T(n, s) \log^2 n + \frac{n^2}{s}\log n)$ expected time using $O(s)$ words of space, for any $s \in \Omega(\log n) \cap O(n)$, in which $T(n, s)$ represents the time complexity of constructing the shortest path tree of a given point inside $P$, with $O(s)$ extra space.






\end{document}